\documentclass[prd,floatfix]{revtex4}

\usepackage{natbib}
\usepackage{graphicx}
\usepackage{xcolor}
\usepackage{graphicx,epsfig}

\def\be{\begin{equation}}
\def\ee{\end{equation}}
\def\ba{\begin{eqnarray}}
\def\ea{\end{eqnarray}}

\begin{document}

\title{Ultra Long-Term Cosmology and Astrophysics}

\author{Robert J. Scherrer\footnote{Email address: 
robert.scherrer@vanderbilt.edu}}
\affiliation{Department of Physics and Astronomy, Vanderbilt University,
Nashville, TN  ~~37235}
\author{Abraham Loeb\footnote{Email address: aloeb@cfa.harvard.edu}}
\affiliation{Harvard-Smithsonian Center for Astrophysics, 60 Garden Street,
Cambridge, MA 02138}
\begin{abstract}

We examine astronomical observations that would be achievable over a future
timeline
corresponding to the documented history of human civilization so far, 
$\sim 10^4$ years.  We examine implications for measurements of the redshift drift,
evolution of the CMB, and cosmic parallax. A number of events that are rare
on the scale of centuries will become easily observable on a timescale
$\sim 10^4$ years.  Implications for several measurements related to gravity are discussed.

\end{abstract}

\maketitle

\section{Introduction}

While most astronomical observations correspond to a ``snapshot" of some object
at a fixed time, there has been increasing interest in longer-term observations.
Perhaps the best current example is the Vera C. Rubin Observatory, which will conduct
a 10-year survey of the entire sky.  However, our ability to utilize data
extending over long (past) timelines is hampered by the relatively short history of astronomy itself.
The era of optical telescopes comprises the past 500 years, with significant
improvements over the past century.  Other branches of astronomy are much more
recent, with radio astronomy nearing its first century of operation, and
observations in other frequencies dating back only to the dawn of the Space Age
and the ability to send telescopes above the Earth's atmosphere (a timescale
which lies well within both authors' lifetimes).  Very long term observations have
been limited to easily measured quantities (e.g., the orbits of the planets).

Now that we are well-situated within the era of modern astronomy, we pose
the following question:  what kinds of observations would be achievable with
an extremely long future timeline?  We need to make an important distinction at this point.
Obviously, we can anticipate technological progress to result in improved instrumentation
in the future, allowing for entirely new windows for observing the universe (as an example, one need look no further than
the recent birth of gravitational wave astronomy).  However, such progress is nearly impossible to predict,
rendering intelligent discussion about future technology extremely difficult.  Instead, we are interested
in observations that are made possible only by a sufficiently long timeline.  For example, a direct measurement
of the rate of visible supernovae within the Galaxy requires that we observe over a sufficiently long time to obtain
a statistically significant number of events; in this case it is the long timeline, and not some supposed
future technology, that allows the measurement to be made.  In what follows, we will assume a current level
of technology extrapolated into the future.  Thus, our claims will represent a lower bound on what could be achieved
with ultra long-term observations.  

What is a reasonable timescale over which future observations are plausible?  The recorded history of human civilization is 
nearly 10,000 years old,
so we will assume that we may extrapolate forward by a similar order of magnitude.  A timescale of 1000 years is too short;
we already have astronomical data going back this far.
Similarly, 100,000 years seems excessively optimistic for the future lifespan of our civilization given that it exceeds recorded 
history and new technologies often bring existential risks.  So we will
base our discussion on an observational timeline from today to 10,000 years in the future.

\section{Cosmology}

The universe is roughly 13.8 Gyr old \cite{Aghanim}, which sets the typical timescale over which significant
changes can be observed in the overall expansion.  Then the typical fractional change in any quantity related to the expansion
observed on a time internal $\Delta t$ will be $\sim \Delta t/(13.8$ Gyr).

The earliest proposal to use long-timeline observations in cosmology is the Sandage-Loeb test \cite{Sandage,Loeb1}.
The Sandage-Loeb test is based on the idea of redshift drift, the change in the redshift $z$ over some time interval
$\Delta t$ due to the expansion of the universe.  Suppose we are measuring the redshift $z$ of an object at some initial
time, and we remeasure this redshift at a later time interval $\Delta t$.  Then
the change in redshift observed over this interval is
\begin{equation}
\Delta z = H_0 \Delta t\left(1+z - \frac{H(z)}{H_0}\right),
\end{equation}
where $H_0$ is the locally-measured value of the Hubble parameter, and $H(z)$ is its value at the redshift $z$.  For $z$ of order
unity, $\Delta z$ is on the order of $\Delta t$ divided by the age of the universe.  A proposed 10-year observation timeline \cite{Loeb1}
yields $\Delta z \sim 10^{-9}$, which is challenging to detect even with the next generation of large 
ground-based telescopes \cite{Geng}
or HI 21-cm observatories \cite{Darling1}.  However, with our assumed timeline of
$\Delta t = 10^4$ yr, we obtain $\Delta z \sim 10^{-6}$, an achievable
sensitivity with existing spectrographs.

The expansion of the universe also causes the cosmic microwave background (CMB) to evolve in time.  The future evolution of
the CMB was first discussed by Lange and Page \cite{Lange} and Zibin, et al. \cite{Zibin}, and the possibility of measuring the evolution
was discussed in detail in Refs. \cite{Moss,Abitbol}.  
Consider first the monopole (i.e., the CMB temperature).  
This temperature, $T$, scales as the inverse of the scale factor $a$, so the time derivative of the temperature is
\begin{equation}
\label{Tevol}
\dot T = - HT.
\end{equation}
Evaluated at the present, we obtain $\dot T = 2 \times 10^{-10}$ K/yr.  Abitbol et al. \cite{Abitbol} examined the plausibility of measuring
this change in $T$ with a ten-year mission and argued that it seems impractical with current technology, largely 
because of foreground
interference.  While a $10^4$ yr timeline produces a relatively large $\Delta T$ ($\sim 10^{-6} $ K), it also opens up the possibility
of nontrivial variation in the foregrounds due to evolution of the Galaxy.

Moss et al. \cite{Moss} argue that time variation of the higher order multipoles
would require $\Delta t \approx 4000$ yr, well within the timeline proposed here,
while the change in the dipole would be detectable over a much shorter timescale.  Note
that the former is due to the cosmological evolution of the CMB itself, while the latter arises from the peculiar velocity
of the solar system motion relative to the zero-dipole frame of the CMB.
However, Loeb \cite{Loeb2} pointed out that the time variation of the quadrupole also receives a contribution from the motion
of the solar system.  This change in the quadrupole moment $Q$ is
at the level of
\begin{equation}
\frac{\dot Q}{Q}\approx 10^{-9} {\rm yr^{-1}}, 
\end{equation}
an order of magnitude larger than 
expected from the intrinsic evolution of the CMB.  Again, this would be easily detectable on the $10^4$ yr time frame considered here.
(See also Ref. \cite{Quercellini} for a review of some of the topics in this section and the following one.)

\section{Cosmic parallax}

Annual parallax, caused by the Earth's orbit about the Sun, provided the
earliest method to directly measure the distances to the nearest stars.  In
addition to this motion, the solar system is moving with respect to the CMB
frame at $369 \pm 0.9$ km s$^{-1}$. (As noted in the previous section,
this motion will eventually result in detectable variations in the CMB). Over a sufficiently long timescale, this
``secular" parallax will allow the measurement of distances to quite distant
objects.  This possibility was first suggested by Kardashev \cite{Kardashev},
and has been examined more recently by several others \cite{Ding,Hall,Paine,Croft,Ferree}.
This parallax shift is given by
\begin{equation}
|\pi| = 77.8 r^{-1} |\sin \beta| \mu {\rm arcsec~yr^{-1}~Mpc}.
\end{equation}
where $r$ is the distance to the object and $\beta$ is the angle between
the direction to the object and the direction of motion of the solar system.
The application of secular parallax to nearby galaxies could provide a measurement
of the Hubble parameter $H_0$ \cite{Hall,Paine,Croft,Ferree}, while quasar parallaxes
would constrain the dark energy equation of state \cite{Ding}.  All of these discussions
assume an observational timeline of 1-10 years.

Now consider
what is possible with an observation time of $10^4$ years.  In that
case, with $\sin \beta \sim 1$, we get a parallax angle of
\begin{equation}
\theta \sim 0.8 r^{-1} {\rm arcsec~Mpc}.
\end{equation}
A galaxy like Andromeda (M31) at a distance of 0.75 Mpc \cite{Riess} would show a parallax on the order
of the
annual parallax of Proxima Centauri.  The distances to low-redshift
galaxies would be easily measurable on this timescale and could establish a robust
calibration of the cosmological distance ladder. 
However,
a major error is introduced by the peculiar velocities
of the galaxies themselves, which cannot simply be removed with a large
sample because of coherent galaxy flows.  Ironically, this source of error also
scales up with observation time.  Techniques to correct for these peculiar
velocities using a large sample of galaxies are discussed in Refs. \cite{Hall,Paine,Croft,Ferree}.
In the case of quasars, the sparse sampling reduces the effects of coherence, and it has been argued
that the proper motion errors will decrease significantly as a function of time \cite{Ding}.

This leads naturally to the possible measurement of transverse velocities using a very long
time baseline \cite{Darling2}, which can be used to confirm the isotropy of the cosmic expansion beyond transverse speeds
induced by large-scale structure and could potentially measure the real-time evolution of the baryon acoustic 
oscillations that serve as a standard ruler in cosmology \cite{Darling3}. At a redshift of $z\sim 0.5$, 
the proper motion expected from cosmic expansion
for the first acoustic peak is $-0.012$ arcseconds per $10^4$ years, which should be measurable on the 
corresponding angular scale of $4.5^\circ$. Even for static objects, the apparent fractional 
compression is 15 $\mu$as per year in the local universe, 
implying a measurable change of 0.15 arcseconds over $10^4$ years.

\section{Rare Events}
Events that are rare on the scale of centuries can be measured with statistical significance on the scale of $10^4$ years,
while events so rare that they have not yet been observed might become observationally accessible on this timescale.
Among the former are supernovae in our galaxy, for which we currently we have a poor estimate of the rate.
The number of
visible supernovae within the past 1000 years is half a dozen, but most supernovae in
our galaxy would have been obscured.  However, we can now observe the neutrinos
from any core collapse supernova in the Galaxy directly.  This rate has
been estimated at $1-2$ supernovae per century.  So over the next 10,000 years,
we would expect to observe $100-200$ core collapse supernovae, allowing for an
estimate of the total rate with an error of about 10\%.

The next 10,000 years are likely to see several potentially destructive astronomical events.
Over this time period, we would
expect to observe $\sim 1$ meteorite impact with a 100 m diameter,
corresponding to an impact energy of 100 megatons of TNT, along with $\sim 10$
Tunguska sized objects \cite{Brown}.  Solar flares with energy on the order of $10^{34}$ erg
have been predicted to occur on the order of once every 2000 years \cite{Shibayama}, so we would expect to observe $\sim 5$ such
events over the next 10,000 years.  Note that these events would have an energy roughly 100 times
as large as the Carrington Event \cite{Tsurutani}.  (See Ref. \cite{Lingam} for a more detailed discussion.)
For both meteorite impacts and solar flares, the probability of an extinction-level event over the next 10,000 years is extremely
small.

Over $10^4$ years, it should also be possible to observe a massive Wolf-Rayet star, like 
Apep (2XMM J160050.7–51424), collapse to a black hole, 
and produce a gamma-ray burst (most likely not in our direction) 
within our own galaxy \cite{Callingham}.  Finally, within the next 10,000 years,
it will be possible to observe three transits of Mercury that coincide with a total solar
eclipse, the soonest in the year 9966, while a transit of Venus
simultaneous with a total solar eclipse will take place just outside of our
fiducial time frame (year 15,232) \cite{Meeus}.

\section{Gravity}
The best current bounds on the time variation of Newton's
constant come from ranging measurements to Mars and the Moon \cite{Will}.
The Mars data give
$\dot G/G = 0.1 \pm 1.6 \times 10^{-13}$ yr$^{-1}$ \cite{Konopliv}, while the limit from the Moon is
$\dot G/G = 4 \pm 9 \times 10^{-13}$ yr$^{-1}$ \cite{Williams1}.
While one might expect these limits to scale up with observation time, this is not
the case for the Mars limit.  The reason is that the orbit of Mars does not constrain
$G$ directly; instead it puts an upper bound on the change in $G M_\odot$.
The current limit is close to the estimates of solar mass loss from light and the solar wind
\cite{Konopliv}.  Thus, any ultra long-term observations of the Mars ephemeris will
not constrain the change in $G$ but will instead provide precision estimates of solar mass loss.
The
Moon, of course, is moving further away from the Earth as it steals our angular
momentum, but this effect can be disentangled from the effect of changing $G$ \cite{Williams2}.
The accuracy of these limits scales as the square of the time span \cite{Williams1}, which
currently corresponds to several decades.  Increasing this by a factor of $\sim 1000$ will
significantly improve the bound on $\dot G/G$.

Constraints on the stochastic gravitational wave background at low frequencies can be steadily improved 
over time with Pulsar Timing arrays \cite{Verbiest}, 
as well as using extragalactic proper motion \cite{Darling3}.  The signal-to-noise for gravitational
waves at a fixed frequency
will scale as $t^{1/2}$ \cite{Siemens}, although this does not take into account further improvements from
adding additional pulsars to the array.

A 10,000 year time span will also allow better constraints on the proper motion of SgrA* \cite{Reid}
and its closest stars \cite{Gualandris}, 
improving current limits on a possible binary companion, invisible stars and stellar remnants, 
and dark matter near the Galactic center.

\section{Discussion}
Astronomy and cosmology with an ultra-long timescale will allow a multitude of observations and measurements that are intrinsically
limited by our current relatively short timescale for modern astronomy.  We do not pretend that the discussion presented here is exhaustive; we suspect that
there are potentially many other observations that such a long timescale would enable.  A related question, not addressed here,
is the issue of maintaining scientific records over a 10,000 year period.  Such problems have been examined
in connection with nuclear waste storage \cite{Hora} and related issues \cite{Benford}.  We also do not consider
the possibility of civilizational collapse, in which case
our descendants will have more important matters to attend to than astronomy.

\section*{Acknowledgments}
We thank S. Taylor and G. Benford for helpful discussions.
A.L. was supported in part by the Black Hole Initiative, which is funded by a grant from JTF and GBMF.
R.J.S. was supported in part by the Department of Energy
(DE-SC0019207).  The authors have no competing interests to declare that are
relevant to the content of this article.  Data sharing is not applicable as
no datasets were generated or analyzed during the current study.

\end{document}